# Anisotropies and magnetic phase transitions in insulating antiferromagnets determined by a Spin-Hall magnetoresistance probe

R. Lebrun[1], A. Ross[1,2,], O. Gomonay[1], S. A. Bender[3], L. Baldrati[1], F. Kronast[4], A. Qaiumzadeh[5], J. Sinova[1], A. Brataas[5], R. A. Duine[3,5,6], & M. Kläui[1,2,5]

[1] Institute for Physics, Johannes Gutenberg-University Mainz, 55099 Mainz, Germany
[2] Graduate School of Excellence Materials Science in Mainz, Staudingerweg 9, 55128, Mainz, Germany
[3] Utrecht University, Princetonplein 5, 3584 CC Utrecht, Netherlands
[4] Helmholtz-Zentrum Berlin für Materialien und Energie, Albert-Einstein Str. 15, 12489, Berlin, Germany
[5] Center for Quantum Spintronics, Department of Physics, Norwegian University of Science and Technology, NO-7491 Trondheim, Norway
[6] Department of Applied Physics, Eindhoven University of Technology, P.O. Box 513, 5600 MB Eindhoven, The Netherlands

**We demonstrate that we can determine the antiferromagnetic anisotropies and the bulk Dzyaloshinskii-Moriya fields of the insulating iron oxide hematite, α-$Fe_2O_3$, using a surface sensitive spin-Hall magnetoresistance (SMR) technique. We develop an analytical model that in combination with SMR measurements, allow for the identification of the material parameters of this prototypical antiferromagnet over a wide range of temperatures and magnetic field values. Using devices with different orientations, we demonstrate that the SMR response strongly depends on the direction of the charge current with respect to the magneto-crystalline anisotropies axis. We show that we can extract the anisotropies over a wide temperature range including across the Morin phase transition. We observe that the electrical response is dominated by the orientation of the antiferromagnetic Néel order parameter, rather than by the emergent weak magnetic moment. Our results highlight that the surface sensitivity of the SMR allows accessing the magnetic anisotropies of antiferromagnetic crystals and in particular thin films where other methods to determine anisotropies such as bulk-sensitive magnetic susceptibility measurements do not provide sufficient sensitivity.**

Antiferromagnets possess a number of intriguing and promising properties for electronic devices, which include a vanishing net moment and thus insensitivity to large magnetic fields [1] and a characteristic terahertz frequency dynamics [2]. However, antiferromagnets are challenging to probe. Since the pioneering work of Louis Néel [1], they have remained the subject of fundamental studies that have mainly relied on synchrotron based facilities for measurements [3]. In recent years, various new effects were discovered which enable more easy and efficient ways to probe and manipulate the antiferromagnetic (or Néel) vector by electrical current [4–6]. The Néel vector can be manipulated by electrical fields in magnetoelectric materials like $Cr_2O_3$ [7] or multiferroics like $BiFeO_3$ [8], by bulk spin-galvanic effects in conducting antiferromagnets [9,10], or by interfacial spin-orbit torques in multilayers with the insulating NiO [11–13]. For magneto-transport measurements, effects that are even functions of the magnetic order parameter like anisotropic [14,15] and spin-Hall magnetoresistance (SMR) [16–18] could probe the antiferromagnetic state and detect switching events [11–13]. However, it is not obvious how one can extract the equilibrium state of an antiferromagnet and in particular determine from the field dependence of the SMR key magnetic properties such as the anisotropy values that are otherwise difficult to ascertain.

The SMR technique can probe the magnetic state of bilayer systems consisting of a ferromagnetic or an antiferromagnetic insulator and a heavy metal. In the simple models [15,22,28], the longitudinal SMR signal $\Delta R$ is proportional to $(1 - (\mathbf{m} \cdot \boldsymbol{\mu})^2)$ for a ferromagnet, and $(1 - (\mathbf{n} \cdot \boldsymbol{\mu})^2)$ for an antiferromagnet (where the unit vector $\hat{\boldsymbol{\mu}}$ of spin-accumulation is perpendicular to the charge current **J** in the heavy metal, **m** and **n** are the magnetization and the Néel vector, respectively). As the magnetization in ferromagnets aligns along the external magnetic field, while the Néel vector of an antiferromagnet tends to align perpendicularly, ferro- and antiferromagnets should show a different field dependence of a SMR signal in the external magnetic field. When the magnetic field is applied parallel to the current, the SMR contribution to the longitudinal resistance should increase for a ferromagnet (positive SMR) and should decrease for antiferromagnets (negative SMR). However, recent experiments show a more complicated behavior of SMR in antiferromagnets, with positive SMR in $SrMnO_3$ [19], negative SMR in the easy-plane NiO [20,21], and both positive [22,23] and negative [24] SMR measured in the easy-axis antiferromagnet $Cr_2O_3$. To explain the different SMR signs, there were suggestions of mechanisms such as contributions from proximity induced magnetization [24] or canted moments [25]. Additional complications stem from the possible strong dependence of the SMR response on the magnetocrystalline anisotropy resulting from structural symmetries [26]. However, such effects, can be concealed in multi-domain states [20,27] by the presence of magneto-elastic coupling.

Hematite, α-$Fe_2O_3$, is not only the main component of rust, but also a prototypical insulating antiferromagnet that lends itself to disentangle the origins of the SMR signals in antiferromagnets and identify the role of easy-axis and easy-plane symmetries in the magnetoresistance response. This iron oxide has a hexagonal crystal structure $R\bar{3}c$, and at room temperature shows an easy-plane canted antiferromagnetic ordering in the basal crystallographic plane. It undergoes a magnetic phase transition at the Morin temperature [28–30] (at about 260 K), below which it exhibits an easy-axis antiferromagnetic ordering along the c-axis. It also possesses a bulk Dzyaloshinskii-Moriya internal field [31,32] along the **c** axis, which is thus hidden in the easy-axis phase and which leads to small (less than a millirad) canting angle between the magnetic sublattices, generating a weak moment in the easy-plane phase. Furthermore, hematite shows an accessible spin-flop field below 10 Tesla so that the Néel vector fully reorients perpendicular to external magnetic fields [33]. The size of the magnetic domains can exceed ten micrometers in both single crystals (see Supplemental) and thin films [34] enabling long distance spin transport [35], and domain redistribution in the easy-plane can be dominated by the coherent rotation of the Néel vector instead of domain wall motion. To understand the spin structures and the switching mechanisms, knowledge of the anisotropies and DMI fields is crucial, which however is challenging in antiferromagnets using conventional approaches requiring large scale facilities.

In this paper, we demonstrate that SMR measurements on R-cut oriented devices of bulk hematite [28,30] can be directly related to the equilibrium orientation of the Néel vector **n** calculated by minimizing the magnetic energy of the sample. By fitting the measured angular dependencies of SMR with theoretically predicted curves we determine the phenomenological constants which define magnetic anisotropies and bulk Dzyaloshinskii-Moriya interactions. We electrically detect the Morin transition through the temperature dependence of the SMR response. In addition, we conclude that the small net magnetization (induced by $H_{DMI}$ or by the external magnetic field **H**) does not significantly contribute to the electrical response in the canted easy-plane phase. These results demonstrate that the SMR technique is a powerful and alternative tool to the complex neutron scattering [3] or X-ray [32] based measurements to ascertain not only the orientation of the Néel vector as demonstrated previously, but also to extract key parameters such as the magnetic anisotropies and DMI fields in antiferromagnets.

We first develop the necessary model that allows us to analyze the results of SMR transport experiments. We assume that the SMR signal depends only on the components of the Néel vector [22, 28]:

$$\frac{\Delta R_{jk}}{R_{jk}} = \rho_0 \begin{cases} 1 - (\mathbf{n} \cdot \mathbf{\mu})^2, & j = k, \\ n_j n_k, & j \neq k, \end{cases} \quad (1.)$$

where $j,k=x,y$, are the coordinates along and perpendicular to the current direction within the sample plane (**Fig. 1. a** and **Fig. 3.a**), the constant $\rho_0$ depends on the spin-mixing conductance at Pt-hematite interface and is considered as a fitting parameter. The unit vector of spin-accumulation **μ** lays within the *xy* plane and is perpendicular to the current. The equilibrium orientation of the Néel vector **n** is calculated by minimizing the potential energy (per unit volume)

$$w_{pot} = 2M_s \left[\frac{1}{2}H_{ex}\mathbf{m}^2 + H_{DMI}(n_X m_Y - n_Y m_X) - \mathbf{H} \cdot \mathbf{m}\right] + w_{ani} \quad (2.)$$

where $H_{ex}$ parametrizes the exchange field which keeps the magnetic sublattices antiparallel, $H_{DMI}$ is the Dzyaloshinskii-Moriya field (DMI), $M_s$ is the sublattice magnetization, **n** and **m** are the Néel vector and magnetization of the antiferromagnet related by the conditions $\mathbf{n} \cdot \mathbf{m} = 0$ and $\mathbf{n}^2 + \mathbf{m}^2 = 1$. The coordinate frame *XYZ* is related to the crystallographic axes and differs from the *xyz* frame defined by the surface plane of the R-cut hematite sample (***x*** being the projection of the easy axis **Z** in the sample plane, ***y*** the axis perpendicular to it) and its normal *z*. The anisotropy energy of the antiferromagnet depends on two parameters:

$$w_{ani} = 2M_s \left[-\frac{1}{2}H_{2\parallel}(T)n_Z^2 - \frac{1}{6}H_\perp(n_X^2 - n_Y^2)(4(n_X^2 - n_Y^2)^2 - 3(n_X^2 + n_Y^2)^2)\right] \quad (3.)$$

The temperature-dependent uniaxial anisotropy field $H_{2\parallel}(T)$ is positive at low temperature, where it stabilizes the easy-axis phase with **n**||**Z**, and changes sign at the Morin temperature. The in-plane anisotropy field $H_\perp > 0$ selects one of three equivalent easy axes in the easy-plane phase as represented in **Fig. 4.c**. Basing on the previously estimated values [28,32,33,36] of the parameters, we assume that $H_{ex} \gg H_{DMI} \gg H_{2\parallel}, H_\perp$, and, hence, $m \ll n \approx 1$.

The external magnetic field induces a rotation of the Néel vector towards a direction perpendicular to the magnetic field. In the easy-axis phase ($H_{2\parallel} > 0$), the final state with $\mathbf{n} \perp \mathbf{H}$ can be achieved either by a spin-flop or by smooth reorientation, depending on the orientation of the magnetic field with respect to the easy axis. A spin-flop takes place at a critical value $H_{sf} = \sqrt{H_{ex}H_{an}^{eff}}$ when **H** is parallel to the easy-axis and competes with the effective anisotropy field $H_{an}^{eff} = H_{2\parallel} - H_{DMI}^2/H_{ex}$. If, in contrast, **H** is perpendicular to the easy-axis, the DMI field pulls the magnetization along the applied field and simultaneously induces a smooth rotation of the Néel vector into the state with $\mathbf{n} \perp \mathbf{H}$ and $\mathbf{n} \perp \mathbf{Z}$. In this case, the final state is achieved at the critical field $H_{DMI,sf} = H_{sf}^2/H_{DMI}$. For a generic configuration with **H** forming an angle $\chi_H$ with the easy axis, the critical field required for a reorientation is

$$H_{cr}(\chi_H) = \frac{H_{DMI,sf}^2 - H_{sf}^2}{2H_{DMI,sf}}|\sin \chi_H| + \frac{1}{2H_{DMI,sf}}\sqrt{(H_{DMI,sf}^2 - H_{sf}^2)^2 \sin^2 \chi_H + 4H_{DMI,sf}^2 H_{sf}^2} \quad (4.)$$

By fitting the angular dependence of the critical field with Eq.(4), we can then determine the parameters $H_{sf}$ and $H_{DMI,sf}$, and calculate the anisotropy and DMI fields, as it will be discussed below.

Experimentally, we access the antiferromagnetic properties of hematite using Hall bar devices (**Fig. 1. a**), aligned along the projection of the c-axis in the R-plane (***x***-axis) or perpendicular to it (***y***-axis). To determine the role of the crystal anisotropies in the SMR response, we first perform measurements at *T*=200 K below the Morin temperature ($T_M = 260$ K) in the easy-axis phase with zero magnetization (**m**=0). We use the two Hall cross geometries, with the devices either perpendicular (***y***-axis) or parallel (***x***-axis) to the easy-axis projection on the sample plane (see sketches in **Fig. 1.a** and **3.a**) and measure the SMR signal by rotating the magnetic field in the sample plane. We define the SMR zero signal to be at zero field.

For devices directed along **y**, we measure the field-dependence of the SMR signal when applying a magnetic field in three directions: parallel (**x** and **y** directions) and perpendicular (**z** direction) to the sample plane, as shown in Fig. **1.b**. When the magnetic field is applied in the sample plane perpendicular to the current J$_y$, **H** ∥ **x**, the SMR signal increases monotonically until it reaches the critical field $H_{cr}(33°)$ =6 T, which corresponds to the transition into the "spin-flop" state with **n** ⊥ **H**. In this state the Néel vector is parallel to the current and does not contribute to the SMR signal, which reaches its maximal value. When the applied magnetic field is perpendicular to the easy axis, **H** ∥ **y**, the SMR signal shows a non-monotonic field dependence, which can be interpreted as a rotation of the Néel vector in the *xz* plane according to theoretical predictions (solid line in **Fig. 1.b**). In this configuration, the magnetic field induces a small magnetization along **y** and the equilibrium orientation of the Néel vector is defined by the competition of the DMI and the easy-axis anisotropy. Above the critical field $H_{cr}(90°) = H_{DMI,sf} \approx 10$ T, the Néel vector reaches its final state perpendicular to the easy axis (and almost perpendicular to the sample plane) with a minor projection on the direction of spin accumulation **μ**. This is seen as the saturation of the SMR signal. Note that we can interpret the experimental data using only the model of pure SMR [17,18], without the need to resort to any proximity effect-induced anisotropic magnetoresistance [15], which would not lead to a change of magnetoresistance for this magnetic transition in the plane perpendicular to the applied current. Finally, for **H** ∥ **z**, we observe monotonic field dependence with an intermediate value of the critical field $H_{cr}(57°)$ =8 T, as expected for an angle of 57° between the applied field and the easy-axis. The agreement between the theoretical model (1)-(3) and experiment indicates that the main contribution to the SMR signal originates from the Néel vector **n** and the contribution of the small induced magnetic moment **m** to the SMR is not significant.

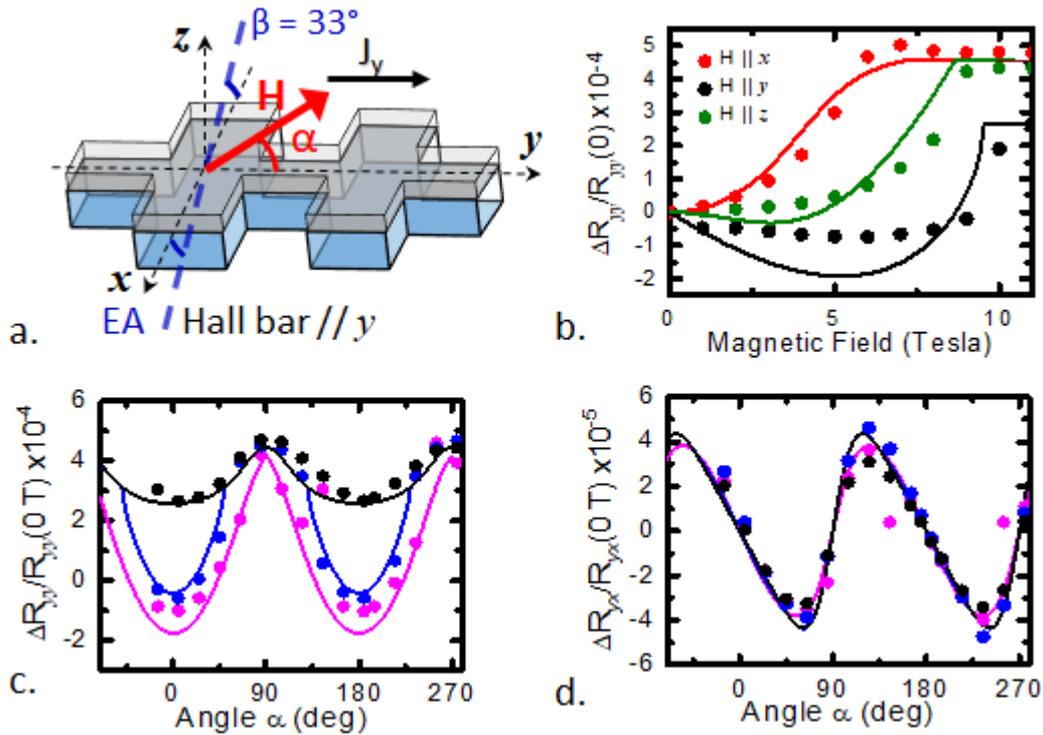

**Figure 1** Spin-Hall magnetoresistance for Pt Hall bars deposited along the *y*-axis (direction of the injected current) of the R-cut single crystal of hematite. (a) Schematic of the devices (b) The longitudinal SMR response as a function of the magnetic fields applied along *x* (red), *y* (black) and *z* (green) axes (c-d) Angular dependence of the longitudinal (c) and transverse (d) SMR for different field values 6 (magenta), 8 (blue) and 11 Teslas (black). The magnetic field rotates within *xy* plane, α is inplane angle between J$_y$ and H. Sample temperature is T=200 K. Theoretical curves (solid lines) are calculated based on the model (1)-(3) with $H_{an}^{eff}$ =23.8 mT, $H_\perp$ = 1.54 μT, $H_{DMI}$ =2.72 T, and $H_{ex}$ = 1040 T [36]. The sample temperature is 200 K.

Figures **1. c-d**, show the angular dependences of the longitudinal and transverse SMR signal measured for the fields around the minimal critical field (6 T), above the maximal critical field (11 T), and also for an intermediate field value (8 T). The magnetic field is rotated within the sample plane, the in-plane angle α is measured relative to the current direction **J_y**, as shown in **Fig. 1.a**. In all cases, the longitudinal SMR signal shows a negative sign (i.e a maximum resistance is achieved for the magnetic field applied perpendicular to the injecting current). The amplitude of the SMR angular dependence peaks at the spin-flop field (**H** = 6 T), when the Néel vector spans all possible states within the sample plane *xy*. In contrast, for the rotation at **H**=11 T, the Néel vector varies between two "spin-flop" states (along *X* and *Y* crystallographic axes) with minor projections on spin-accumulation axis *x*. This explains the lower value of SMR signal. The transverse SMR signal (**Fig. 1.d**) shows similar behavior with a maximal amplitude at 6 T.

Reasonable agreement between experimental data and theoretical modelling for all angular and field dependencies of SMR signal shows that our single-domain approximation is sufficient to describe the system and allows us to estimate the values for the DMI constant and $H_{an}^{eff}$ based on the minimal and maximal values of critical fields (in combination with the reported value of the exchange field ($H_{ex}$=1040 T [36]). However, these parameters can be determined directly with higher accuracy from fitting the angular dependence of the critical field **H_cr** measured by rotating the magnetic field in the *xz* plane, see **Fig. 2**.

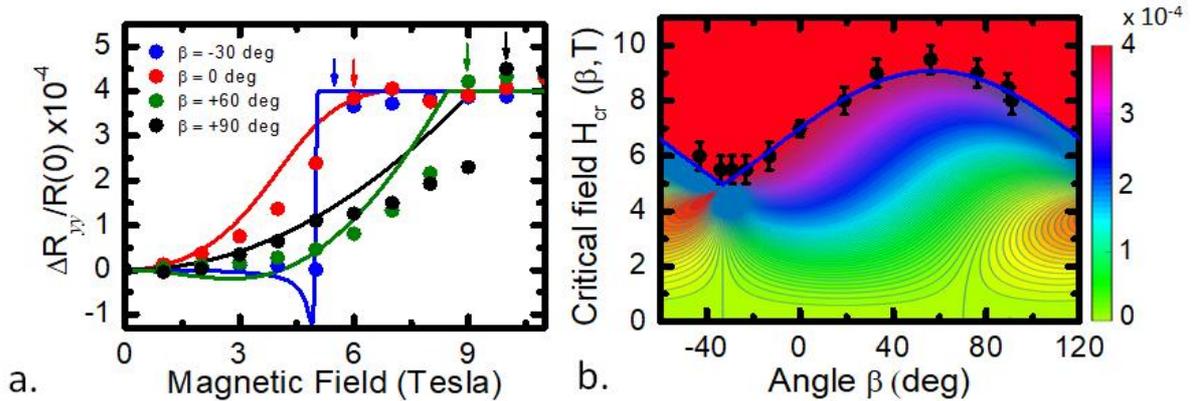

**Figure 2** Longitudinal Spin-Hall magnetoresistance in the β-plane of Pt Hall bars deposited along the *y*-axis. (a) Field dependence of the longitudinal SMR signal for magnetic fields applied parallel (β = –30°) and perpendicular (β = 60°) to easy axis and for two intermediate orientations in *xz* plane (β = 0° and 90°). (b) Dependence of the longitudinal SMR (color code) as a function of the amplitude and direction of the magnetic field in *xz* plane calculated from the model (1)-(3) with $H_{an}^{eff}$ =23.8 mT, $H_\perp$ = 1.54 μT, and $H_{DMI}$ =2.72 T. Solid line (theory) and dots (experiment) show the angular dependence of the critical field. Vertical lines correspond to the field scans in the panel. Vertical arrows correspond to the extracted critical fields. (a). Angle β is calculated from *x* axis, current is along *y*, as shown in Fig. 1.a. In calculations we account for shift between crystallographic and experimental axes, $\chi_H = (\beta + 33)°$. The sample temperature is T=200 K.

**Figure 2** shows the results of such measurements for a Pt Hall bar directed along *y*. In **Fig. 2.a**. we demonstrate typical field dependences of the longitudinal SMR for different orientations of the magnetic field. When the magnetic field is parallel to the easy axis (β = -30°), the SMR curve changes step-wise, thus reflecting the spin-flop reorientation of the Néel vector. Such an abrupt transition is typical for easy-axis antiferromagnets, which can possess only indistinguishable 180° domains. This behavior is in contrasts with the smooth variation of the SMR signal in easy-plane antiferromagnets, where the magnetic field can induce domain wall motion between non-180° domains [20,27]. For other field directions, the SMR signal varies smoothly, in accordance with theoretical predictions (solid lines). The critical field in these cases is associated with the points of saturation of the SMR signal (vertical arrows).

In **Fig. 2.b.** we present the full field-angular dependence of the longitudinal SMR signal calculated according to Eqs.(1)-(3). The black solid line delineates the region of the spin-flop state (with **n** ⊥ **H**) and the corresponding saturation of SMR signal. This line, determined from Eq. (4), coincides with the experimental angular dependence of the critical field $H_{cr}(\chi_H)$. By fitting the experimental data $H_{cr}(\beta)$ (dots) with the predicted dependence $H_{cr}(\chi_H)$ (taking into account that $\chi_H = \beta + 33°$), we extract the values of the internal DMI and effective anisotropy fields as $H_{\text{DMI}}$ = 2.72 T and $H_{an}^{eff}$ =23.8 mT. These values are consistent with the previous measurements [37] and show that the simple surface-sensitive electrical measurements allow for a direct and precise access to the state and magnetic properties of antiferromagnetic system.

In antiferromagnets, the SMR response strongly depends on the orientation of the anisotropy axis relatively to the patterned devices. Contrary to ferromagnets with low anisotropies like YIG or NiFe, the projection of the Neel vector along the spin-accumulation can even be zero as long as the applied magnetic fields are lower than the large critical fields. To identify and corroborate the role of anisotropy fields in the transport measurements, we thus repeat the measurements for Pt Hall bars rotated by 90°, with injected currents parallel to the *x*-axis (**Fig. 3.a**). For this geometry of the devices, the Néel vector is perpendicular to the spin accumulation at zero field. The field dependencies of longitudinal SMR measured for **H**∥*x* and **H**∥*z* (**Fig. 3.b**. are in good agreement with the theoretical model (Eqs. (1)-(3)) and show the same critical fields (6 and 8 T correspondingly) as the devices with a *y*-oriented electrode. For **H**∥*y* the SMR signal is nearly constant, as the Néel vector rotates in the plane perpendicular to the spin-accumulation. In this case, the SMR response does not reflect the Néel reorientation observed for Pt stripes directed along **y**. The small quantitative discrepancy between the theoretical model and the experiments at high magnetic fields, larger than the critical fields, could be associated with a residual contribution from the emerging canting moment and from the limitations of our analytical model. The angular dependences of the longitudinal (**Fig. 3.c**) and transverse (**Fig. 3.d**) SMR are negative, in agreement with the model predictions (Eq. 1). This further confirms that we indeed measure a pure SMR effect with no significant proximity induced magnetic moment contributions. Note that in spite of a visual difference in the SMR dependencies measured for *x*- and *y*-oriented devices, all the data can be consistently interpreted within the same theoretical model that we develop. This also means that SMR signal must be analyzed in regard to the device orientation relative to the crystallographic axis. In our case, information from the *x*-oriented devices is limited, due to the absence of an electrical signature of the DMI induced reorientation for **H**∥*y*.

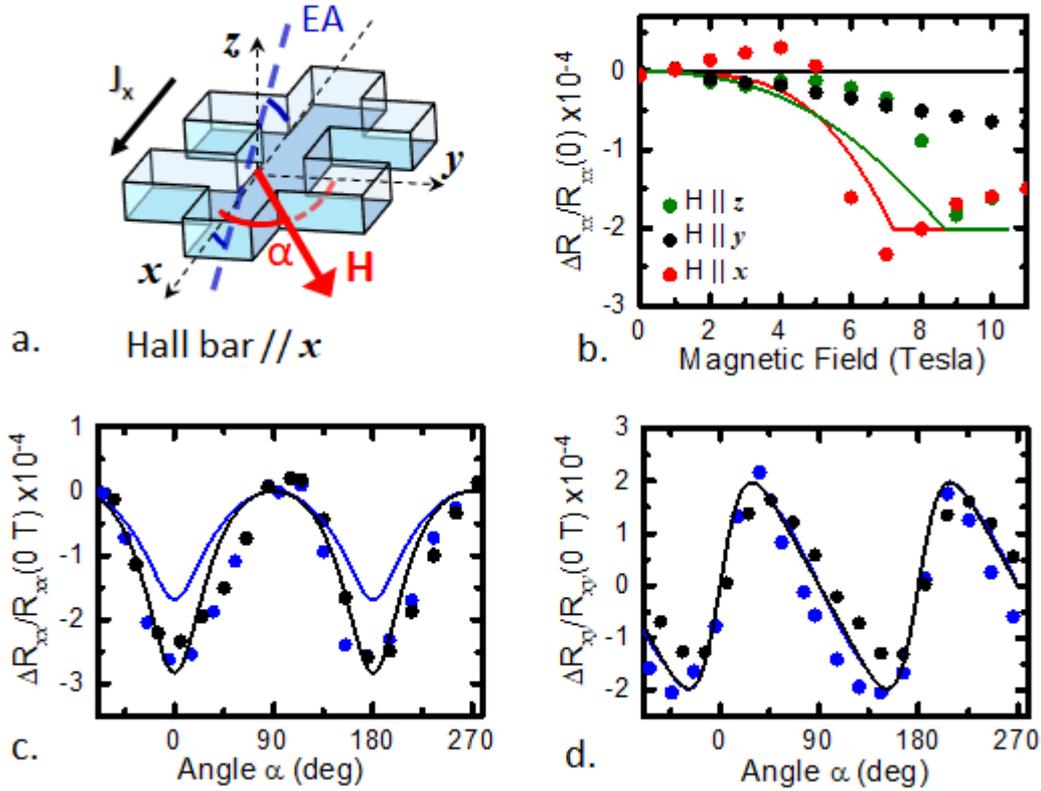

**Figure 3** Spin-Hall magnetoresistance for Pt Hall bars deposited along the *x*-axis (direction of the injected current) of the R-cut single crystal of hematite. (a) Schematic of the devices, α is an angle between $J_x$ and H. (b) The longitudinal SMR response as a function of the magnetic fields applied along *x* (red), *y* (black) and *z* (green) axes (c-d) Angular dependence of the longitudinal (c) and transverse (d) SMR for H = 8 (blue) and 11 T (black).

Finally, the SMR technique allows one to probe the equilibrium states of hematite over a wide temperature range and studying the (*T,H*) magnetic phase diagram. To illustrate this, we measure field dependences of the longitudinal SMR for *x*-oriented devices at different temperatures (see **Fig. 4.a**), below and above the Morin transition temperature which separates the easy-axis and easy-plane phases. Extracted values of the critical fields (closed black dots) are shown in **Fig. 4.b** as a function of temperature. Within the interval 160-255 K the temperature dependence of the critical field is well described by $H_{cr}(T) \propto \sqrt{T_M - T}$, with $T_M = 255$ K. Such a temperature dependence is typical for the second-order phase transitions within the Landau theory and was also reported earlier[1] [36,38]. Below 160 K, $H_{cr}(T)$ stays around a constant value of 8 T, above 255 K its value is close to zero. Using Eq.(4) and experimental dependence $H_{cr}(T)$ we then determine the temperature dependence of the effective anisotropy $H_{an}^{eff}(T)$ (blue line in **Fig.4.b**) assuming that $H_{DMI} = 2.72$ T is temperature-independent (as previously found [37,39]). The last assumption is consistent with the observed (open dots in **Fig. 4.b**) and predicted (solid line) temperature dependence of the maximal critical field[2]. Note, that as the SMR is a surface sensitive technique, similar measurements could determine the magnetic anisotropies of high quality epitaxial hematite thin films [34], for which magnetometry measurements such as SQUID do not provide sufficient sensitivity.

---

[1] We succeeded to fit the reported data from papers [36,38] with the same law.
[2] Due to limitations of our setup we can access the maximal critical field for **H** ⊥ Easy-axis only down to 200 K.

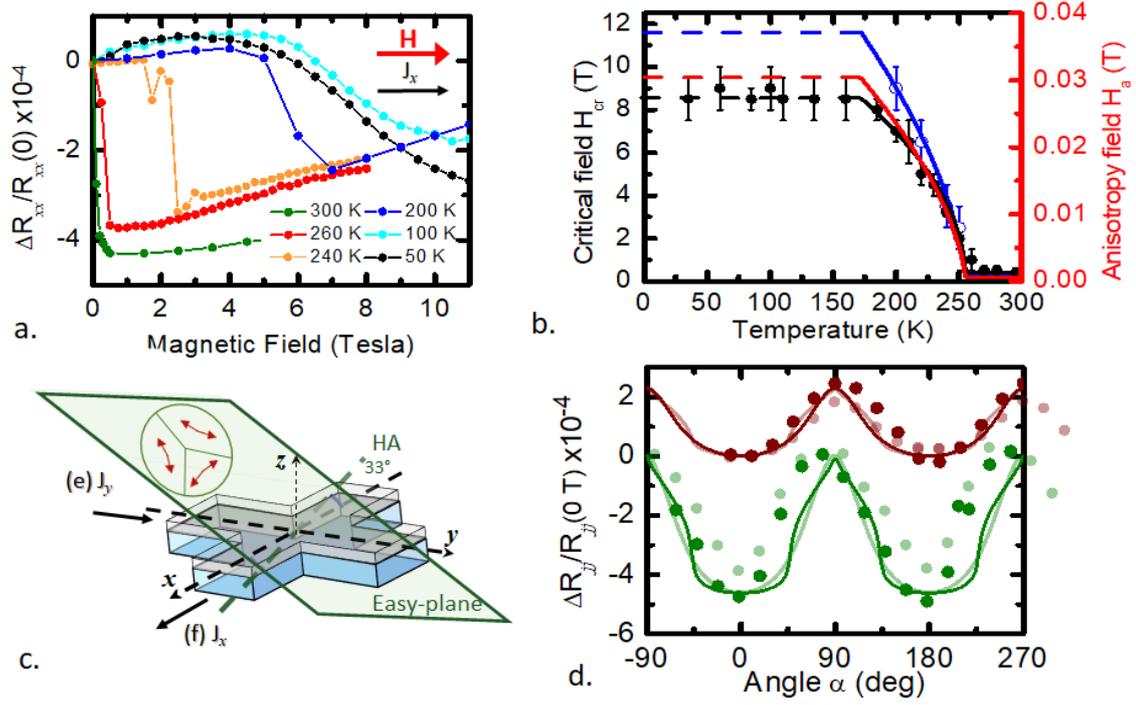

**Figure 4** Temperature dependence of the Spin-Hall magnetoresistance (a) Field dependence of the longitudinal SMR for H parallel to J ($||x$) at different temperatures (below and above the Morin temperature of 260 K). (b) Temperature dependence of the critical field $H_{cr}$ (left axis) for H$||x$ ($H_{cr}(T; 33°)$, black dots) and H $|| y$ ($H_{cr}(T; 90°)$, blue dots). Solid lines shows the fitting with $H_{cr} \propto \sqrt{T_M - T}$. Blue line (right axis) shows the calculated temperature dependence of the effective anisotropy $H_{an}^{eff}(T)$ based on fitting of critical fields and Eq.(4). (c) Schematic of the two device orientations (J$||y$ and J$||x$). Shaded is the easy plane with orientations (double-arrows) of the Neel vectors in different domains. (d) Angular dependence of the longitudinal SMR signal for J$||y$ (green) and J$||x$ (brown) at $H=0.5$ T (full dots) and $H=5.5$ T (open dots) in easy-plane phase ($T=280$ K). Field is rotated in $xy$ plane, angle α is calculated from the current direction. Symbols are experimental data; solid lines correspond to fitting with $H_{an}^{eff} = -23.8$ mT, $H_\perp = 1.54$ µT, and $H_{DMI} = 2.72$ T.

According to the theoretical model, the effective anisotropy vanishes at the Morin temperature, $H_{an}^{eff}(T_M) = 0$, and changes sign and becomes negative in an easy-plane phase (above $T_M$). However, the SMR signal at 260 and 300 K still shows the features of the spin-flop transition at small field around 0.4 T (see **Fig. 4.a**). Moreover, the SMR has the same sign below and above $T_M$ which indicates that the SMR is still dominated by the Néel vector and not by the small canted moment. Assuming that reorientation of the Néel vector in the easy-plane phase is driven mainly by in-plane anisotropy (see Eq. (3)), we estimate the value $H_\perp$ from the angular dependencies of SMR measured for small (0.5 T) and large (5.5 T) fields applied in $xy$ plane and for the two orientations of Pt stripes (**Fig. 4.d**). As seen from **Fig. 4.d**, the amplitude of the SMR signal remains the same for small and large fields. According to theoretical predictions this means that the critical field in the easy-plane phase is below 0.4 T, which gives $H_\perp \leq 1.6$ µT. The weak field-dependence of the SMR amplitude also means that the field-induced "weak" magnetic moment only contributes negligibly. Note, that for geometrical reasons, the SMR amplitude is larger for $x$-oriented device demonstrating again the role of the crystal and device symmetry in the electrical response.

**To conclude, we have combined theory with experiments to analyze the properties of the insulating antiferromagnetic hematite using accessible transport measurements. We successfully model SMR measurements in α-Fe$_2$O$_3$ by a simple analytical model. From comparison and fitting of our electrical measurement results, we determine the values of anisotropy and DMI fields, which drive the Néel vector reorientation. Measuring across the Morin transition, we determine the specific properties of the easy-axis and canted easy-plane phases, and observe that the**

contribution of the weak moments is negligible for the electrical response, which is governed by the Néel vector. We demonstrate that the shape of the SMR curves strongly depends on the orientation of the devices with respect to the crystallographic axis in line with our model predictions. Our observations demonstrate that SMR technique is a unique surface sensitive technique that allows us to easily access to the fundamental properties of insulating antiferromagnets with the need to resort to large scale facility measurements as previously required.

## Methods

**Lithography:**

The devices were carried out based on a sample geometry that was defined using electron beam lithography and the subsequent deposition and lift-off of a 7 nm platinum layer by DC sputtering in an argon atmosphere at a pressure of 0.01 mbar. The devices were contacted using a bilayer of chromium (6 nm) and gold (32 nm). The sample was mounted to a piezo-rotating element in a variable temperature insert that was installed in a superconducting magnet capable of fields up to 12 T and cooled with liquid helium. For the rotation measurements, the sample was rotated in a constant field.

**Single crystal of α-Fe$_2$O$_3$:**

The single crystal of hematite, α-Fe$_2$O$_3$, was obtained commercially and is orientated with [1120] out of plane which means that the c-axis of the hexagonal structure is tilted 34 degrees below the surface plane. This orientation was chosen because of the large in-plane projection of the c-axis in addition to the stability of the sample terminating with the r-plane.

**Data availability statement:**

The data that support the findings of this study are available from the corresponding authors upon reasonable request. Correspondence and requests for materials should be addressed to R.L. or M.K (e-mails: rxlebrun@gmail.com or klaeui@uni-mainz.de).